\documentclass[prl,twocolumn,showpacs,amsmath,amssymb,floatfix,superscriptaddress]{revtex4-1}

\usepackage{graphicx}%Include figure files
\usepackage{dcolumn}%Align table columns on decimal point
\usepackage{bm}% bold math

\begin{document}

\title{Spin Hall Effect and Origins of Nonlocal Resistance in Adatom-Decorated Graphene}

\author{D. Van Tuan}
\affiliation{Catalan Institute of Nanoscience and Nanotechnology (ICN2), CSIC and The Barcelona Institute of Science and Technology, Campus UAB, Bellaterra, 08193 Barcelona, Spain}
\affiliation{Department of Electrical and Computer Engineering,
University of Rochester, Rochester, NY 14627, USA.}
\author{J. M. Marmolejo-Tejada}
\affiliation{Department of Physics and Astronomy, University of Delaware, Newark, DE 19716-2570, USA}
\affiliation{School of Electrical and Electronics Engineering, Universidad del Valle, Cali, AA 25360, Colombia}
\author{X. Waintal} 
\affiliation{CEA-INAC/UJF Grenoble 1, SPSMS UMR-E 9001, Grenoble F-38054, France}
\author{B. K. Nikoli\'{c}}
\affiliation{Department of Physics and Astronomy, University of Delaware, Newark, DE 19716-2570, USA}
\author{S.O. Valenzuela}
\affiliation{Catalan Institute of Nanoscience and Nanotechnology (ICN2), CSIC and The Barcelona Institute of Science and Technology, Campus UAB, Bellaterra, 08193 Barcelona, Spain}
\affiliation{ICREA--Institucio Catalana de Recerca i Estudis Avan\c{c}ats, 08010 Barcelona, Spain}
\author{S. Roche}
%\email{stephan.roche@icn.cat}
\affiliation{Catalan Institute of Nanoscience and Nanotechnology (ICN2), CSIC and The Barcelona Institute of Science and Technology, Campus UAB, Bellaterra, 08193 Barcelona, Spain}
\affiliation{ICREA--Institucio Catalana de Recerca i Estudis Avan\c{c}ats, 08010 Barcelona, Spain}

\date{\today}
\begin{abstract}
Recent experiments on the spin Hall effect (SHE) in graphene with adatoms, reporting unexpectedly large charge-to-spin conversion efficiency, have raised a fierce controversy. Here,  we apply numerically exact Kubo and Landauer-B\"{u}ttiker formulas to realistic disordered models of gold-decorated graphene (including adatom segregation) to compute the spin Hall conductivity and spin Hall angle, as well as the nonlocal resistance which is directly measured in experiments. Large spin Hall angles of $\simeq 0.1$ are obtained at zero-temperature, but their dependence on adatom segregation differ from the predictions of semiclassical transport theories. Furthermore, our findings evidence multiple contributions to the nonlocal resistance, some unrelated to the SHE, and a strong suppression of the SHE at room temperature, which altogether cast doubts on recent claims of giant SHE in graphene. All this calls for future experiments to unambiguously reveal the existence of SHE physics and clarify the upper limit on spin current generation by two-dimensional materials.
\end{abstract} 

\maketitle

\textit{Introduction}.{---}Over the past decade, the spin Hall effect (SHE) has evolved rapidly from an obscure theoretical prediction to a major resource for spintronics ~\cite{Vignale2010,Sinova2015}. In the direct SHE, injection of conventional unpolarized charge current into a material with extrinsic (due to impurities) or intrinsic (due to band structure) spin-orbit coupling (SOC) generates pure spin current in the direction transverse to charge current. Although SHE was first observed only a decade ago \cite{Kato2004}, it is already ubiquitous within spintronics as standard pure spin-current generator and detector \cite{Vignale2010,Sinova2015}. The ratio of the spin Hall current to the charge current, i.e., spin Hall angle ($\theta_\mathrm{sH}$) is the figure of merit of charge to spin current conversion efficiency. To date, measured values of $\theta_\mathrm{sH}$ range from $\sim 10^{-4}$ for semiconductors to $\sim 0.1$ for metals such as Au, Ta, Pt or \mbox{$\beta$-W} \cite{Sinova2015}. 

On the other side, the discovery of graphene \cite{Neto2009} has been followed by a considerable amount of activity, owing to its unique electronic properties and versatility for practical applications, including spintronics \cite{Ferrari2015,Roche2015}. The intrinsically small SOC and hyperfine interactions in graphene lead to spin relaxation lengths reaching several tens of micrometers at room temperature \cite{Hernando2006,Tombros2007,Han2010,Seneor2012,Guimarães2014}, but simultaneously making clean graphene inactive for SHE. Recently, by decorating graphene with heavy metal adatoms like copper, gold, silver, and fluorine,  Balakrishnan and coworkers have reported one of the largest values of $\theta_\mathrm{sH}\simeq 0.2-0.9$ ever measured \cite{Balakrishnan2014}.  
These reports follow prior work on weakly hydrogenated graphene, which surprisingly showed similar results~\cite{Balakrishnan2013}. These large values of $\theta_\mathrm{sH}$ have been supported by semiclassical transport analysis \cite{Ferreira2014,Yang2016}. 

However, contradictory findings were obtained experimentally for both hydrogenated and functionalized (with metallic adatoms) graphene \cite{Wang2015a,Kaverzin2015}. There, the presence of large nonlocal signals was confirmed but appeared to be disconnected from SHE physics.  Wang and coworkers \cite{Wang2015a} reported that Au- or Ir- decorated graphene exhibit no signature of SHE, and relate the large nonlocal resistance ($R_\mathrm{NL}$) to the formation of neutral Hall currents, whereas Kaverzin and van Wees found large $R_\mathrm{NL}$ signal in hydrogenated graphene, however insensitive to an applied in-plane magnetic field \cite{Kaverzin2015}. The authors exclude valley Hall effect and long-range chargeless valley currents as mediating such $R_\mathrm{NL}$, given the absence of both temperature dependence and broken inversion symmetry \cite{Gorbachev2014,Ju2015,Kirczenow2015,Ando2015}, and conclude that a nontrivial and unknown phenomenon is at play.

Additionally, the theoretical interpretation of a largely enhanced SOC (estimated about 2.5 meV) by hydrogen defects~\cite{Balakrishnan2013} is contradicted by {\it ab-initio} calculations \cite{Gmitra2013a}, which predicts SOC one order of magnitude smaller, and very short spin lifetime \cite{Kochan2014}. Besides, the presently available theories for $\theta_\mathrm{sH}$~\cite{Ferreira2014} or $R_\mathrm{NL}$~\cite{Abanin2009} offer little guidance on how to resolve these controversies since they utilize semiclassical approaches to charge transport and spin relaxation which are known to break down near the charge neutrality point (CNP)~\cite{Tuan2014,Chen2012a}. Moreover, while Kubo formula~\cite{Streda1982} offers fully quantum-mechanical treatment that can in principle capture all relevant effects, its standard analytical evaluation neglects~\cite{Sinitsyn2006} terms in the perturbative expansion in disorder strength which can become crucial for clusters of adatoms (such as those corresponding to skew-scattering from pairs of closely spaced impurities~\cite{Ado2015}).  Finally the impact of unavoidable adatom clustering \cite{Sutter2011} onto graphene surface on $\theta_\mathrm{sH}$ is an open and important question, since adatom segregation has been shown to strongly affect spin transport properties \cite{Pi2009,Cresti2014}.

In this Letter, the spin Hall angles in graphene decorated by Au-adatoms are computed by using numerically exact quantum transport theory, namely the Kubo transport method and the multiterminal Landauer-B\"{u}ttiker (LB) scattering approach.  At zero temperature, both methods yield $\theta_\mathrm{sH}\sim 0.1-0.3$ for the same Au-adatom density. However, those values, which require a large density of adatom coverage (above $10\%$), drop significantly when temperature and adatom clustering are taken into account.  Additionally, a finite contribution to $R_\mathrm{NL}$ (with large negative sign) is obtained when SOC is artificially turned off, suggesting possible non-SHE contributions to large $R_\mathrm{NL}$ signals, already observed in quasiballistic regimes in different materials \cite{Mihajlovic, Wang2016}.

\textit{Hamiltonian model for Au-decorated graphene}.{---}When an adatom like gold, thallium or indium is absorbed onto graphene surface, it places in the hollow sites of graphene carbon rings and induces both the intrinsic type  $V_I$ and Rashba SOC $V_R$ on 
 carbon atoms in these rings due to the broken inversion symmetry \cite{Weeks2011}. The $\pi$-$\pi$* orthogonal tight-binding model for graphene under the effect of such kind of adatoms reads
 \begin{eqnarray}
{\mathcal{H}}=&-&\gamma_0\sum_{\langle ij\rangle }c_i^+c_j+\frac{2i}{\sqrt{3}}V_{I}\sum_{\langle\langle ij\rangle\rangle \in \mathcal{R}}c_i^+\vec{s}\cdot(\vec{d}_{kj}\times\vec{d}_{ik})c_j\nonumber\\
&+&iV_R\sum_{\langle ij\rangle \in \mathcal{R}}c_i^+\vec{z}\cdot(\vec{s}\times\vec{d}_{ij})c_j-\mu\sum_{i\in \mathcal{R}}c_i^+c_i.
\label{Hamil}
\end{eqnarray}
The first term is the nearest neighbor hopping term with $\gamma_0 = 2.7$ eV. The second term is a next-nearest neighbor hopping term which represents the intrinsic SOC induced by the adatoms. $\vec{d}_{kj}$ is the unit vector pointing from atom $i$ to atom $k$, with atom $k$ standing in between $i$ and $j$, and $\vec{s}$ is a vector defined by the Pauli matrices $(s_{x},s_{y},s_{z})$. The third term describes the Rashba SOC which explicitly violates $\vec{z} \rightarrow -\vec{z}$ symmetry. The last term is the potential shift $\mu$ associated to the carbon atoms in the hexagonal rings $\mathcal{R}$ containing adatoms, simulating some charge modulation induced very locally around the adatom \cite{Weeks2011}. Such model has been employed to study spin dynamics in graphene \cite{Tuan2014}.  Here we take the same parameters $V_{I}=0.007\gamma_{0}, V_{R}=0.0165\gamma_{0} $ and $\mu=0.1\gamma_{0}$ derived from {\it ab-initio} calculations \cite{Weeks2011}.

\begin{figure}
\includegraphics[scale=0.15,angle=0]{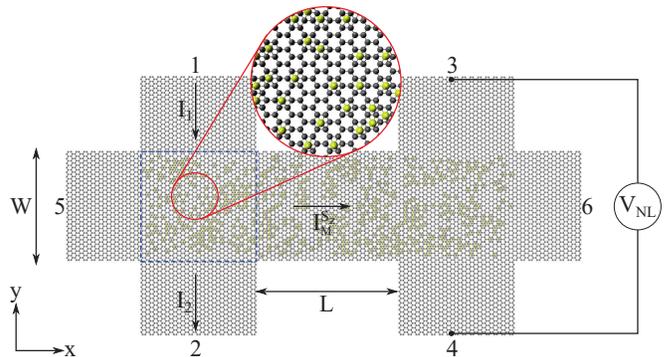}
\caption{(Color online) Schematic view of a six-terminal graphene geometry used for computing the nonlocal resistance $R_\mathrm{NL}=V_\mathrm{NL}/I_1$ and spin Hall angle $\theta_\mathrm{sH}$ determined as $\theta_\mathrm{sH}=I_{5}^{s_{z}}/I_{1}$.  For nonlocal transport, the injected transverse charge current between electrodes 1 and 2 generates the longitudinal spin current $I_5^s$ in lead 5, as well as the putative mediative spin current $I^s$ whose conversion into the voltage drop $V_\mathrm{NL}$, between the electrodes 3 and 4, leads to $R_\mathrm{NL}$. The dashed region denotes the sample size \mbox{$400$ nm $\times$ $400$ nm}, with periodic boundary conditions used for calculations of Kubo conductivities. In the zoom, black circles represent carbon atoms and yellow circles label positions of scattered or clustered Au adatoms.}
\label{fig1}
\end{figure}

Figure \ref{fig1} shows the geometry used for the calculations of bulk Kubo conductivities and multiterminal charge and spin currents. The calculations of $\theta_\mathrm{sH}$ with the Kubo formula are performed using a graphene region of the size \mbox{$400$ nm $\times$ $400$ nm} enclosed in dashed square (with periodic boundary conditions). For LB calculations of $R_\mathrm{NL}=(V_3-V_4)/I_1$ and $\theta_\mathrm{sH}=I^{s_{z}}_5/I_1$, we consider a full six-terminal bridge of width \mbox{$W=50$ nm} and with variable distance $L$ between the pair of electrodes 1 and 2 and the pair of electrodes 3 and 4 (see Fig. \ref{fig1}). Such geometry and the determination of $\theta_{\mathrm{sH}}$ mimics the experimental procedure used in Refs.~\cite{Balakrishnan2013,Balakrishnan2014,Wang2015a,Kaverzin2015} 

\textit{Calculation of the Spin Hall conductivity and dissipative conductivity}.{---}The Kubo formula for spin Hall conductivity $\sigma_\mathrm{sH}$ reads \cite{Sinova2015} 
 \begin{equation}
 \sigma_\mathrm{sH}=\frac{e\hbar}{\Omega}\sum_{m,n}\frac{f(E_m)-f(E_n)}{E_m-E_n}\frac{\mathcal{I}m[\left\langle m\left|J_x^z\right|n \right\rangle \left\langle n\left|   v_y\right|m \right\rangle]}{E_m-E_n+i\eta},
\label{EQSC}
 \end{equation}
where $J_x^z=\frac{\hbar}{4}\{s_z,v_x\}$  is the spin current operator, $s_z$ the z-component of the Pauli matrix. The calculation of Eq. (\ref{EQSC}) is usually made by computing the whole spectrum $E_m$ and the full set of eigenvectors $\{|m\rangle\}$ of ${\mathcal{H}}$ which is a highly expensive numerical task. Here we develop alternatively an efficient real-space formalism by rewritting $\sigma_\mathrm{sH}$ as
 \begin{equation}
 \sigma_\mathrm{sH}
 =\frac{e\hbar}{\Omega}\int dxdy\frac{f(x)-f(y)}{(x-y)^2+\eta^2}j(x,y),
 \end{equation}
with $j(x,y)=\sum_{m,n}\mathcal{I}m[\left\langle m\left|J_x^z\right|n \right\rangle \left\langle n\left|   v_y\right|m \right\rangle] \delta(x-E_m) \delta(y-E_n)$, which can be calculated by rescaling $\mathcal{H}, x,y$ and $ E$ into $[-1,1]$ (the corresponding variables are $\hat{H}, \hat{x},\hat{y}$ and $\hat{E}$ respectively) and by using 2D-expansion of $j(x,y)$ in Chebyshev polynomials $T_m(\hat{x})$ as $j(x,y)=\sum_{m,n}^M (4\mu_{mn}g_mg_nT_m(\hat{x})T_n(\hat{y}))/((1+\delta_{m,0})(1+\delta_{n,0})\pi^2\sqrt{(1-\hat{x}^2)(1-\hat{y}^2)})$, where $\mu_{mn}=\frac{4}{\Delta E^2}\mathcal{I}m[Tr[J^z_xT_{n}(\hat{H})v_yT_{m}(\hat{H})]]$, and $\Delta E$ is the bandwidth \cite{Weisse2006}. Here $g_m$ is the filter, Jackson kernel, that minimizes the Gibbs oscillations arising in truncating the series to finite order $M$ \cite{Weisse2006}. The trace in $\mu_{mn}$ is computed by the average on a small number $R\ll N$ of random phase vectors $|\varphi\rangle$.  Hereafter $M=1500$ (resp. 6000) for $\sigma_\mathrm{sH}$ (resp. $\sigma_{xx}$), $R=1$, $N=4\times 10^6$. Similar real-space methods have been developed for the longitudinal conductivity $\sigma_{xx}$ \cite{Ferreira2015}, Hall conductivity $\sigma_{xy}$ \cite{Ortmann2014,Garcia2015} and spin Hall conductivity $\sigma_\mathrm{sH}$ \cite{Berg2011}. The method is here validated for both intrinsic and Rashba-SOC clean cases by comparison with prior studies in the quantum spin Hall regime and with analytical derivations \cite{Dyrdal2009} (see Supplemental information \cite{DinhSupp}).

\begin{figure}[htbp]
 \includegraphics[width=0.45\textwidth]{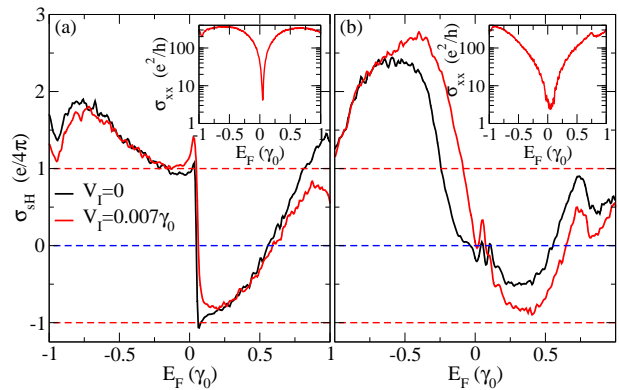}
\caption{(color online): $\sigma_\mathrm{sH}$ (main frame) and $\sigma_{xx}$ (insets) for two cases of 15\% Au-adatoms distributions: Scattered (left panel) and clustered distributions of Au into islands of varying radius in $[1,3]$ nm (right panel). In both cases the effect of presence (red lines) or absence (black lines) of intrinsic SOC is shown. Results are averaged over $400$ disorder realizations.}
 \label{fig2}
 \end{figure}

\textit{Spin Hall angles for various adatom distributions}.{---} Figure \ref{fig2} shows $\sigma_\mathrm{sH}$ for $15\%$ of Au adatoms distributed in scattered distributions (Fig. \ref{fig2}(a)) and clustered coverages (Fig. \ref{fig2}(b)) defined by randomly distributed islands of radius \mbox{$\in [1,3]$ nm}. Although the random distribution of Au impurities (and related Rashba-SOC field) induce scattering [$\mu=0.1\gamma_0$ in Eq. (\ref{Hamil})], in absence of intrinsic SOC, the energy profile of $\sigma_\mathrm{sH}$ is reminiscent from the typical step behavior obtained for an homogeneous Rashba field \cite{DinhSupp}, with a characteristic value of  $\sigma_\mathrm{sH}\simeq \pm e/4\pi$ near half filling. Adding a small intrinsic SOC ($V_I=0.007\gamma_{0}\ll V_R$)  slightly changes the absolute value of $\sigma_\mathrm{sH}$  but preserves the step behavior. In contrast, the clustered distribution of Au adatoms suppresses the step behavior and smoothen the shape of $\sigma_\mathrm{sH}$ close to CNP, a fact demonstrating that clustering effectively decreases the effective Rashba SOC. The effect of intrinsic SOC is more pronounced for the clustered distribution with a more significant enhancement of $\sigma_\mathrm{sH}$ on both electron and hole sides. 

\begin{figure}[htbp]
\includegraphics[scale=0.3,angle=0]{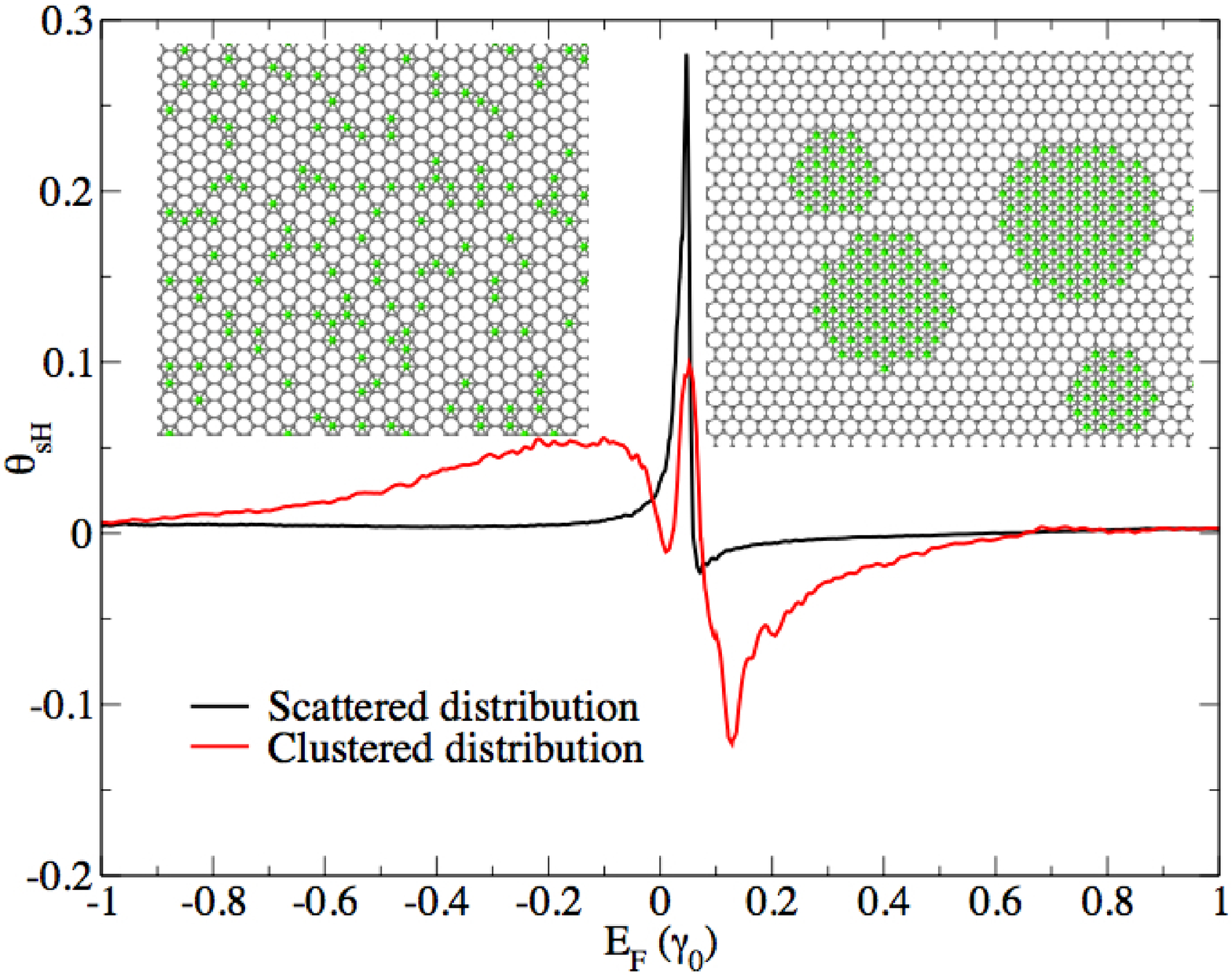}
\caption{(color online) Spin Hall angle $\theta_\mathrm{sH}$ for two cases of 15\% gold adatoms distribution: Scattered (in black) and clustered distributions (in red), illustrated in insets.}    
\label{fig3}
\end{figure}

The spin Hall angle $\theta_\mathrm{sH}=\sigma_\mathrm{sH}/\sigma_{xx}$ also requires the calculation of the longitudinal conductivity $\sigma_{xx}$, which is performed using a real space Kubo approach \cite{DinhSupp}. Figure \ref{fig2} (insets) shows $\sigma_{xx}$ for both cases. Comparable values of $\sigma_{xx}$ are obtained at CNP, but for the scattered case, $\sigma_{xx}$ increases with energy faster than for the clustered case. Figure \ref{fig3} shows $\theta_\mathrm{sH}$ for 15\% of Au adatoms on the graphene surface, which are distributed homogeneously (black lines) or in clusters (red line). Remarkably, $\theta_\mathrm{sH}$ shown in Fig. \ref{fig3} are very large, in the order of few $0.1$ close to CNP, which is similar to the experimental values \cite{Balakrishnan2014}. However a threefold decrease in $\theta_\mathrm{sH}$ is obtained when adatoms are segregated into islands with small radius, manifesting the detrimental effect of adatom clustering on SHE. This contradicts the semiclassical transport predictions where $\theta_\mathrm{sH}$ increases with the radius of adatoms clusters \cite{Ferreira2014}. Moreover, our calculations indicate a downscaling of $\theta_\mathrm{sH}$ with the adatom density ($n_{i}$) as $\theta_\mathrm{sH}\sim n_{i}$, which leads to much smaller $\theta_\mathrm{sH}\sim 10^{-4}-10^{-5}$ for the adatom density estimated in the experiments \cite{Balakrishnan2014} (see Supplemental Information). 

\textit{Nonlocal charge transport and SHE}.{---}In the SHE experiments \cite{Balakrishnan2013,Balakrishnan2014},  a multiterminal graphene transport configuration is used as illustrated in Fig.~\ref{fig1}. In such a circuit, a charge current $I_1$ injected from lead 1 towards lead 2 generates a nonlocal resistance $R_\mathrm{NL}=(V_3 - V_4)/I_1$ for the Fermi energy $E_F$ tuned within the some interval around the CNP. The appearance of non-zero $R_\mathrm{NL}$, due to a SHE-driven mechanism, is explained by charge current $I_1$ inducing spin current $I^S$ in the first crossbar in Fig.~\ref{fig1} flowing in the direction $5 \rightarrow 6$, which is subsequently converted into the nonlocal voltage $V_\mathrm{NL}=V_3 - V_4$  by the inverse SHE in the second crossbar. 

We calculate the total charge and spin currents via the LB formulas (implemented in {\tt KWANT}~\cite{Groth2014}) using the above multiterminal geometry to directly compare our results with the experimental work. Then we calculate the nonlocal resistance and investigate its origin and its association with the SHE.  Computationally, $I_p = \sum_q G_{p q}(V_p - V_q)$ connects the total charge current $I_p$, flowing through semi-infinite ideal (i.e., without impurities or SOC)  graphene lead \mbox{$p=$1--6} attached to the central region decorated with Au adatoms, with voltages $V_q$ in all other leads (see Fig. \ref{fig1}).  Spin currents $I_p^{s_\alpha}$ are derived using spin-dependent transmission coefficients \cite{DinhSupp}. One then applies the equation for $I_p$ to the circuit in Fig.~\ref{fig1} by either setting the voltages $V_q$ to find currents $I_p$ and $I_p^{s_\alpha}$, or fix the injected current $I_p$ and then find voltages $V_q$ which develop as the response to it.  For comparison with experiments, we consider $I_1$, the current injected through lead 1 and current $-I_1$ flows through lead $2$ while $I_p \equiv 0$ in the other four leads. We then compute voltages which develop in the leads \mbox{$p=$3--6}  in response to injected current $I_1$ and obtain the nonlocal resistance $R_\mathrm{NL}$. 

\begin{figure}
\includegraphics[scale=0.38,angle=0]{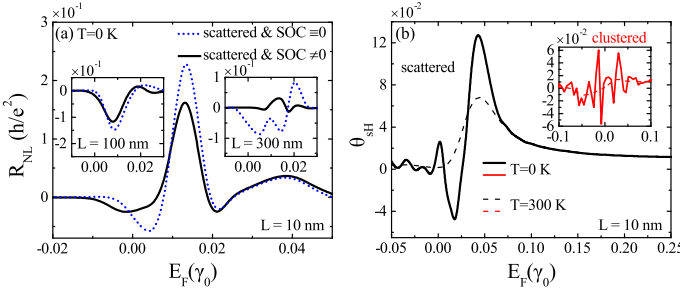}
\caption{(Color online) (a) Nonlocal resistances $R_\mathrm{NL}$ as a function of the Fermi energy ($E_{\rm F}$) for various channel lengths: $L=10$ nm (main frame); $L=100$ nm (left inset), $L=300$ nm (right inset) and channel width $W=50$ nm (Fig.~\ref{fig1}) for 15\% of scattered Au adatoms.  $R_\mathrm{NL}$ of non-SHE (SOC $\equiv 0 \Leftrightarrow  V_I=V_R=0$) origin are shown in dotted lines. (b) $\theta_\mathrm{sH}$ for the same adatom coverage for scattered (main frame) and clustered (inset) adatoms, at T=0 K and T=300 K. All curves are averaged over 10 configurations.}
\label{fig4}
\end{figure}
We first determine the spin Hall angle with LB by $\theta_\mathrm{sH}=I_{5}^{s_{z}}/I_{1}$, and compare it with the obtained results derived from the Kubo formula. As shown in Fig.~\ref{fig4} (b), LB calculations confirm the large values of $\theta_\mathrm{sH}$ obtained by the Kubo formula, as well as the detrimental effect of clustering of Au adatoms which significantly reduces $\theta_\mathrm{sH}$, by at least a factor of 2 for the chosen cluster geometry. While both Kubo and LB calculations predict $\theta_\mathrm{sH}\simeq 0.1$ close to the CNP, thermal broadening effects obtained in LB calculations further reduce $\theta_\mathrm{sH}$ by up to one order of magnitude [see Fig.~\ref{fig4}(b)]. By comparing Fig. \ref{fig4}(b) with Fig. S3 in \cite{DinhSupp}, we find that the hypothetical case of an homogeneous SOC, with Au adatoms covering uniformly every hexagon in Fig. \ref{fig1},  generates an intrinsic SHE with $\theta_\mathrm{sH}$ vs. $E_F$ exhibiting wider peak (centered at 0.3 $\gamma_0$ due to doping of graphene by Au adatoms) of slightly smaller magnitude than in the case of randomly scattered Au adatoms (covering 15\% of hexagons). Thus, adatom induced resonances at the CNP play a crucial role in generating large spin Hall current and nonlocal voltage $R_\mathrm{NL}$.

Fig. \ref{fig4}(a) shows $R_\mathrm{NL}$ as a function of energy for various channel lengths. Notably, we found a non-zero $R_\mathrm{NL}$ even when {\em all SOC-terms are switched off} ($V_R=V_I=0$), while keeping the other disorder potential terms in Eq. (1) unchanged. The change of sign of $R_\mathrm{NL}$ in Fig. 4(a) with increasing channel length from $L=10$ nm to $L=300$ nm suggests the following interpretation---the total $R_\mathrm{NL}$  has in general four contributions $R_\mathrm{NL}=R_\mathrm{NL}^\mathrm{SHE}+R_\mathrm{NL}^\mathrm{\mathrm{Ohm}}+R_\mathrm{NL}^\mathrm{qb}+R_\mathrm{NL}^\mathrm{pd}$  (assuming they are additive after disorder averaging). For unpolarized charge current injected from lead 1 (i.e., electrons injected from lead 2): $R_\mathrm{NL}^\mathrm{SHE}$ due to SHE has a positive sign; trivial Ohmic contribution $R_\mathrm{NL}^\mathrm{Ohm}$ due to classical diffusive charge transport \cite{Abanin2009} has a positive sign; $R_\mathrm{NL}^\mathrm{qb}$ is the negative quasiballistic contribution arising due to direct transmission $T_{32} \neq 0$ from lead 2 to lead 3, as observed in experiments on multiterminal gold Hall bars \cite{Mihajlovic}; finally $R_\mathrm{NL}^\mathrm{pd}$ is a positive contribution  which is specific to Dirac materials where evanescent wavefunctions generates pseudodiffusive transport regime \cite{Tworzydlo2006,Chang2014a} close to CNP characterized by conductance $G \propto 1/L$ even in perfectly clean samples with $W > L$ (see Fig. S4-Supplemental Materials) showing non-zero $R_\mathrm{NL}$ in perfectly clean Hall bars without SOC, as long as $W \ge L$). 

The negative sign of $R_\mathrm{NL}$ in the two insets of Fig. 4(a) for $V_R=V_I=0$ and  $L>W$ suggests that $R_\mathrm{NL}^\mathrm{Ohm}$ can be neglected in our samples due to small concentration of adatoms, so that $R_\mathrm{NL}^\mathrm{qb}$ competes with positive sign $R_\mathrm{NL}^\mathrm{SHE}$. In graphene bars with $W>L$, positive sign $R_\mathrm{NL}$  [such as for $W=50$ nm, $L=10$ nm example in Fig. 4(a)] is dominated by $R_\mathrm{NL}^\mathrm{pd}$. Thus, the existence of contributions to $R_\mathrm{NL}$ which do not originate from SHE and can be much larger than $R_\mathrm{NL}^\mathrm{SHE}$ can explain the insensitivity of the total $R_\mathrm{NL}$ to the applied external in-plane magnetic field observed in some experiments \cite{Wang2015a,Kaverzin2015}.  

Additional spin-sensitive experiments with improved control of functionalization are crucial to establish the strength of SHE in graphene with enhanced SOC. The difficulty in clarifying the dominant contribution to nonlocal resistance could be resolved by detecting its sign change as a function of the Hall bar length. Our analysis indeed suggests that upper limit of $R_\mathrm{NL}^\mathrm{SHE}$ could be isolated and quantified through LB calculations performed on setup where adatoms would be completely removed in the channel of length $L$ in Fig.~1. When such ballistic channel [see illustration in Fig. S5(a)-Supplemental Materials] is sufficiently long, $R_\mathrm{NL}^\mathrm{pd} =0$,  $R_\mathrm{NL}^\mathrm{qb} \rightarrow 0$ due to the absence of impurity scattering in the channel and spin current $I^S$ generated by direct SHE in the first crossbar arrives conserved at the second crossbar where it is converted into $V_\mathrm{NL}$ by the inverse SHE. We observe that the upper limit of $R_\mathrm{NL}^\mathrm{SHE}$ obtained by this procedure in Fig. S6(b)--Supplemental Materials is smaller than the absolute value of other contributions to $R_\mathrm{NL}$ in Fig.~4(a) at intermediate channel lengths. 

\begin{acknowledgments}
This work has received funding from the European Union Seventh Framework Programme under grant agreement 604391 Graphene Flagship. S.R. acknowledges the Spanish Ministry of Economy and Competitiveness for funding (MAT2012-33911), the Secretaria de Universidades e Investigacion del Departamento de Economia y Conocimiento de la Generalidad de Cataluna and the Severo Ochoa Program (MINECO SEV-2013-0295). We acknowledge computational resources from PRACE and the Barcelona Supercomputing Center (Mare Nostrum),
under Project 2015133194.  J.M.M-T. was supported as Fulbright Scholar and by Colciencias (Departamento Administrativo de Ciencia, Tecnologia e Innovacion) of Colombia. B.K.N. was supported by NSF Grant No. ECCS 1509094, and is grateful for the hospitality of Centre de Physique Th\'{e}orique de Grenoble-Alpes, where part of this work was done. The supercomputing time was provided by XSEDE, which is supported by NSF Grant No. ACI-1053575.
\end{acknowledgments}

%********************references************************************************************************

%\bibitem{Dumas2014}
%R.K. Dumas {\it et~al.}, IEEE Trans. Magn.  {\bf 50}, 4100107 (2014).
%\bibitem{Gmitra2009}
%M. Gmitra, S. Konschuh, C. Ertler, C. Ambrosch-Draxl, and J. Fabian, Phys. Rev. B {\bf 80},  235431  %(2009).
%\bibitem{Niimi2015}
%Y. Niimi and Y. Otani, Rep. Prog. Phys. {\bf 78},  124501  (2015).
%\bibitem{Hoffmann2013}
%A. Hoffmann, IEEE Trans. Magn. {\bf 49},  5172  (2013). 
\end{document}